\documentclass[aps,prl,twocolumn,showpacs,floatfix,amsmath,amssymb]{revtex4}

\usepackage{amsmath,bm,amsfonts}
\usepackage{graphicx}

\newfont{\bg}{cmr10 scaled\magstep4}

\newcommand{\bigzerou}{\smash{\lower1.7ex\hbox{\bg 0}}}

\begin{document}

\title{Core level binding energies in solids from first-principles}
  
\author{Taisuke Ozaki and Chi-Cheng Lee}
\affiliation{
    Institute for Solid State Physics, The University of Tokyo, Kashiwa 277-8581, Japan
}

\date{\today}

\begin{abstract} 
A general method is presented to calculate absolute binding energies of core levels in metals and insulators, 
based on a penalty functional and an exact Coulomb cutoff method in a framework of the density functional theory.
The spurious interaction of core holes between supercells is avoided by the exact Coulomb cutoff method,
while the variational penalty functional enables us to treat multiplet splittings due to chemical shift, 
spin-orbit coupling, and exchange interaction on equal footing, both of which are not accessible by previous methods.
It is demonstrated that the absolute binding energies of core levels for both metals and insulators 
are calculated by the proposed method in a mean absolute (percentage) error of 0.4 eV (0.16~\%) for eight cases 
compared to experimental values measured with X-ray photoemission spectroscopy
within a generalized gradient approximation to the exchange-correlation functional. 
 
\end{abstract}


\maketitle

Since the pioneering works by Siegbahn et al. \cite{Siegbahn1967,Siegbahn1969}, the X-ray photoelectron spectroscopy (XPS) 
has become one of the most important and widely used techniques in studying chemical composition 
and electronic states in the vicinity of surfaces of materials \cite{Fadley2010}.
Modern advances combined with synchrotron radiation further extend its usefulness to enable
a wide variety of analyses such as core level satellites \cite{Bagus2010}, core level vibrational fine structure \cite{Hergenhahn2004}, 
magnetic circular dichroism (MCD) \cite{Yang2002}, spin-resolved XPS \cite{Kim2001}, and photoelectron holography \cite{Omori2002}.    
The basic physics behind the still advancing XPS measurements is dated back to the first interpretation
for the photoelectric effect by Einstein \cite{Einstein1905}. 
An incident X-ray photon excites a core electron in a bulk, and the excited electron with
a kinetic energy is emitted from the surface to the vacuum. The binding energy of the core 
level in the bulk can be obtained by the measurement of the kinetic energy \cite{Siegbahn1967,Siegbahn1969}. 
Theoretically, the calculation of the binding energy involving the evaluation of
the total energies for the initial and final states is still a challenging issue 
especially for insulators, since after the emission of the photoelectron the system is 
not periodic anymore and ionized due to the creation of the core hole. 
The violation of the periodicity hampers use of conventional electronic structure methods 
under a periodic boundary condition, and the Coulomb potential of the ionized bulk 
cannot be treated under an assumption of the periodicity due to the Coulombic divergence.
One way to avoid the Coulombic divergence is to neutralize the final state with a core hole 
by adding an excess electron into conduction bands \cite{Pehlke1993,Susi2015,Olovsson2010} 
or to approximate the bulk by a cluster model \cite{Bagus2013}.  
However, the charge compensation may not occur in insulators because of the short escape time 
of photoelectron ($\sim 10^{-16}$ sec.) \cite{Cavalieri2007}, while the treatment might be 
justified for metals.
Even if we employ the charge compensation scheme, the screened core hole pseudopotential 
which has been widely used in pseudopotential methods allows us to calculate only the chemical 
shift of binding energies, but not the absolute values \cite{Pehlke1993}. 
A method based on the Slater transition state theory \cite{Slater1974} for bulks also relies 
on the charge compensation \cite{Olovsson2010}. 
In spite of the long history of the XPS and its importance in materials science, 
a general method has not been developed so far to calculate the absolute binding energies 
for both insulators and metals, including multiplet splittings due to chemical shift, spin-orbit coupling,
and exchange interaction, on equal footing. 
In this Letter we propose a general method to calculate absolute binding energies of core levels 
in metals and insulators, allowing us to treat all the issues mentioned above, 
in a single framework within the density functional theory (DFT) \cite{Hohenberg1964,Kohn1965}. 

\begin{figure}[b]
    \centering
    \includegraphics[width=8.5cm]{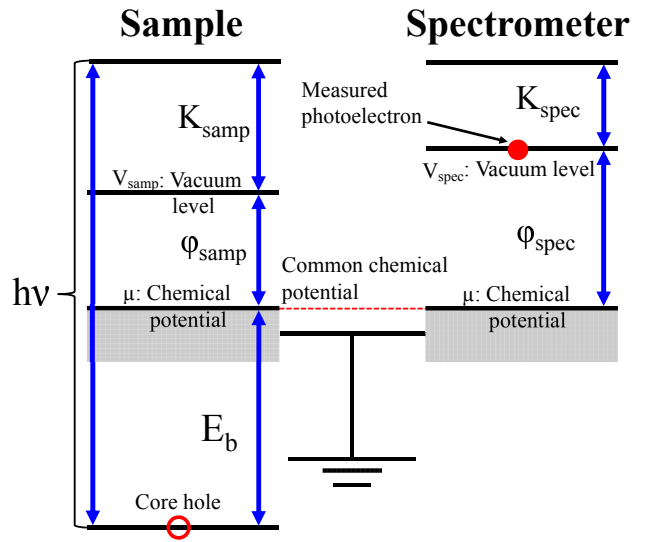}
    \caption{Schematic energy diagram for a sample and a spectrometer in the XPS measurement.
             }
\end{figure}

Let us start to define the absolute binding energy $E_{\rm b}$ of core electrons in bulks measured 
by a XPS experiment, based on the energy conservation. The energy of the initial state is given by 
the sum of the total energy $E_{\rm i}(N)$ of the ground state of $N$ electrons and an energy $h\nu$ of 
a monochromatic X-ray photon. On the other hand, that of the final state is contributed by the total 
energy $E_{\rm f}(N-1)$ of the excited state of $N-1$ electrons with a core hole, and the kinetic energy $K_{\rm spec}$
of the photoelectron placed at the vacuum level $V_{\rm spec}$ as shown in Fig.~1. 
Therefore, the energy conservation in the XPS measurement is expressed by 
\begin{eqnarray}
  E_{\rm i}(N) + h\nu = E_{\rm f}(N-1) + V_{\rm spec} + K_{\rm spec},
\end{eqnarray}
where the left and right hand sides of Eq.~(1) correspond to the initial and final states, respectively.
Noting that the chemical potential of the sample is aligned with that of the spectrometer $\mu$ 
by Ohmic contact, and that the vacuum level of the spectrometer $V_{\rm spec}$ is given by 
$V_{\rm spec}=\mu + \varphi_{\rm spec}$ using the work function of the spectrometer $\varphi_{\rm spec}$,  
Eq. (1) reads to 
\begin{eqnarray}
  h\nu - K_{\rm spec} - \varphi_{\rm spec} = E_{\rm f}(N-1) - E_{\rm i}(N) + \mu.
\end{eqnarray}
The left hand side of Eq.~(2) is what is defined as the binding energy $E^{\rm (bulk)}_{\rm b}$ measured 
by the XPS experiment if the charging of the sample is carefully compensated, 
and the recoil energy is ignored \cite{Watts1994}. 
The right hand side of Eq.~(2) provides a useful expression 
to calculate the absolute binding energy $E^{\rm (bulk)}_{\rm b}$ for bulks regardless of the band gap of materials.
However, the chemical potential $\mu$ during the XPS measurement is not accessible in general, while 
the chemical potential $\mu$ might be calibrated using the Fermi edge of silver as a standard procedure 
in the XPS experiment \cite{King2007}.
Thus, it is desirable to calculate the binding energy $E^{\rm (bulk)}_{\rm b}$ without relying on the experimental 
chemical potential $\mu$. 
One should notice that what the shift of chemical potential $\Delta\mu$ does is only the constant 
shift of potential, and that $\mu$ differs from the {\it intrinsic} chemical potential $\mu_0$ by $\Delta\mu$, 
where {\it intrinsic} means a state which is free from the control of chemical potential. 
Then, one may write
$E_{\rm i}(N)=E^{(0)}_{\rm i}(N)+N\Delta\mu$ 
and 
$E_{\rm f}(N-1)=E^{(0)}_{\rm f}(N-1)+(N-1)\Delta\mu$
using the {\it intrinsic} total energies $E^{(0)}_{\rm i}(N)$ and $E^{(0)}_{\rm f}(N-1)$ 
by assuming the common chemical potential $\mu$ for both the initial and final states 
due to a very large $N$.
Inserting these equations into Eq.~(2) yields
\begin{eqnarray}
  E^{\rm (bulk)}_{\rm b} = E^{(0)}_{\rm f}(N-1) - E^{(0)}_{\rm i}(N) + \mu_0.
\end{eqnarray}
Equation (3) is an important consequence, since only quantities which can be calculated from first-principles
without suffering from details of each experimental condition are involved. Thereby, we use Eq.~(3) to calculate 
the absolute binding energy $E^{\rm (bulk)}_{\rm b}$. 
It should be emphasized that Eq.~(3) is valid even for semi-conductors and insulators. 
In an arbitrary gapped system, the common chemical potential $\mu$ is pinned at either the top of valence band or 
the bottom of conduction band. Or $\mu$ is located in between the top of valence band and the bottom of conduction band. 
For all the possible cases, it is confirmed that the exactly same discussion above is valid. 
Thus, one may conclude that Eq.~(3) is a general formula to calculate 
the absolute binding energy $E^{(\rm bulk)}_{\rm b}$ of solids. 
Especially for metals, Eq.~(3) can be further reorganized by noting a relation derived with the Janak theorem \cite{Janak1978}: 
$E^{(0)}_{\rm f}(N-1)-E^{(0)}_{\rm f}(N)=\int dn \partial E^{(0)}_{\rm f}/\partial n = -\mu_0$, 
where $n$ is an occupation number of an one-particle eigenstate on the Fermi surface, $dn= -ds/S$ defined with the area of 
the Fermi surface $S$ and an infinitesimal area $ds$,  and the surface integral is performed over the Fermi surface. 
By inserting the above equation into Eq.~(3), we obtain the following formula:
\begin{eqnarray}
  E^{\rm (metal)}_{\rm b} = E^{(0)}_{\rm f}(N) - E^{(0)}_{\rm i}(N), 
\end{eqnarray}
which allows us to employ the total energy of the neutralized final state  $E^{(0)}_{\rm f}(N)$
instead of that of the ionized state. For metals, Eqs.~(3) and (4) should result in an equivalent 
binding energy in principle, however, the convergence is different from each other as a function of 
the system size as shown later on.
On the other hand, for gaseous systems the common chemical potential does not exists anymore, and instead both vacuum levels 
of the sample and the spectrometer are aligned to zero by adjusting or calibrating $V_{\rm spec}$ to zero 
using an electrometer or a known ionization potential of a noble gas measured at the same time \cite{Siegbahn1969}.
From Eq.~(1) with $V_{\rm spec}$ of zero, it is found that the absolute binding energy $E^{\rm (gas)}_{\rm b}$ 
is expressed by 
\begin{eqnarray}
  E^{\rm (gas)}_{\rm b} = E^{(0)}_{\rm f}(N-1) - E^{(0)}_{\rm i}(N).
\end{eqnarray}
Comparing Eq.~(5) with Eq.~(3), we see that the two definitions for the binding energy differs by $\mu_0$.
$E^{(0)}_{\rm i}(N)$ and $\mu_0$ in Eqs.~(3) can be calculated by a conventional approach with 
the periodic boundary condition.

So, we now turn to discuss a method of calculating $E^{(0)}_{\rm f}(N-1)$ in Eq.~(3)
based on the total energy calculation including many body effects naturally.
Core electrons for which a core hole is created are explicitly included in the calculations 
to treat multiplet splittings due to chemical shift, spin-orbit coupling, and exchange 
interaction between core and spin-polarized valence electrons, and to take account 
of many body screening effects. The creation of the core hole can be realized by expressing 
the total energy of the final state by the sum of a conventional total energy $E_{\rm DFT}$ 
within DFT and a penalty functional $E_{\rm pen}$ as 
\begin{eqnarray}
  E^{(0)}_{\rm f}(N-1) = E_{\rm DFT} + E_{\rm pen}
\end{eqnarray}
with the definition of $E_{\rm pen}$:
\begin{eqnarray}
  E_{\rm pen} = \frac{1}{V_{\rm B}} \int_{\rm B} dk^3 \sum_{\mu}f_{\mu}^{({\bf k})}
              \langle \psi_{\mu}^{({\bf k})}\vert \hat{P} \vert \psi_{\mu}^{({\bf k})} \rangle,
\end{eqnarray}
where $\int_{\rm B}dk^3$ is the integration over the first Brillouin zone of which volume is $V_{\rm B}$, 
$f_{\mu}^{({\bf k})}$ the Fermi-Dirac function, 
and $\psi_{\mu}^{({\bf k})}$ the Kohn-Sham wave function of two-component spinor.
In Eq.~(7) the projector $\hat{P}$ is defined with 
an angular eigenfunction $\Phi$ of the Dirac equation under a spherical potential 
and a radial eigenfunction $R$ obtained by an atomic DFT calculation 
for the Dirac equation as 
\begin{eqnarray}
  \hat{P} \equiv \vert R\Phi_{J}^{M}\rangle \Delta \langle R\Phi_{J}^{M}\vert
\end{eqnarray}
with for $J=l+\frac{1}{2}$ and $M=m+\frac{1}{2}$
\begin{eqnarray}
  \nonumber
  \vert \Phi_{J}^{M}\rangle 
   &=&
    \left(
     \frac{l+m+1}{2l+1}   
    \right)^\frac{1}{2}
    \vert Y_{l}^{m} \alpha \rangle
   + 
    \left(
     \frac{l-m}{2l+1}   
    \right)^\frac{1}{2}
    \vert Y_{l}^{m+1} \beta \rangle,\\
\end{eqnarray}
and for $J=l-\frac{1}{2}$ and $M=m-\frac{1}{2}$
\begin{eqnarray}
  \nonumber
  \vert \Phi_{J}^{M}\rangle 
   &=&
    \left(
     \frac{l-m+1}{2l+1}   
    \right)^\frac{1}{2}
    \vert Y_{l}^{m-1} \alpha \rangle
   - 
    \left(
     \frac{l+m}{2l+1}   
    \right)^\frac{1}{2}
    \vert Y_{l}^{m} \beta \rangle,\\
\end{eqnarray}
where $Y$ is a spherical harmonic function, and $\alpha$ and $\beta$ spin basis functions.
The variational treatment of Eq.~(6) with respect to $\psi$ leads to the following Kohn-Sham equation:
\begin{eqnarray}
  \left(
    \hat{T} + v_{\rm eff} + \hat{P}  
  \right)\vert \psi_{\mu}^{({\bf k})}\rangle
   = 
  \varepsilon^{({\bf k})}_{\mu}\vert\psi_{\mu}^{({\bf k})}\rangle,
\end{eqnarray}
where $\hat{T}$ is the one-particle kinetic operator, and $v_{\rm eff}$ the conventional KS effective potential 
originated from $E_{\rm DFT}$.
If a large number, 100 Ryd. was used in this study, is assigned for $\Delta$ in Eq.~(8), 
the targeted core state $\Phi_{J}^{M}$ specified by the quantum numbers $J$ and $M$ is penalized
through the projector $\hat{P}$ in Eq.~(11), 
and becomes unoccupied, resulting in the creation of a core hole for the targeted state.
Since the creation of the core hole is self-consistently performed, the screening effects by both 
core and valence electrons, spin-orbit coupling, and exchange interaction are 
naturally taken into account in a single framework. 
It is also straightforward to reduce the projector $\hat{P}$ to the non-relativistic treatment. 
In addition, the variational treatment allows us to perform geometry optimization for the final state 
with the core hole. 

\begin{figure}[t]
    \centering
    \includegraphics[width=8.5cm]{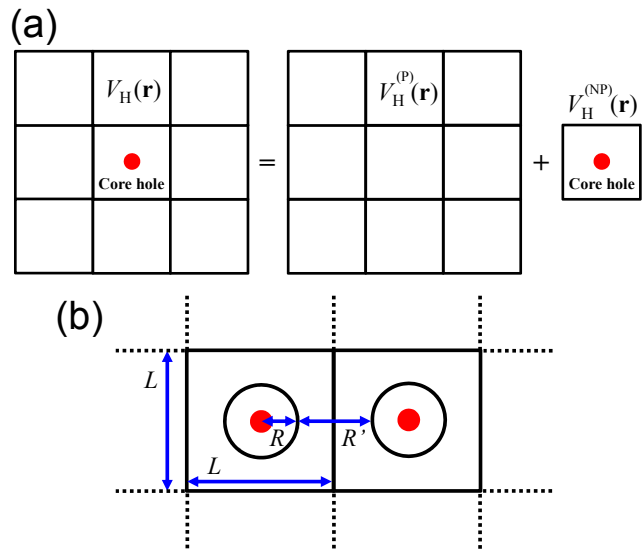}
    \caption{(a) Treatment of the Hartree potential in a system with a core hole under the periodic boundary condition.
             (b) Configuration to calculate 
             the non-periodic part of the Hartree potential $V_{\rm H}^{({\rm NP})}$ 
             by the exact Coulomb cutoff method for $\Delta \rho$. 
             }
\end{figure}

After the creation of the core hole, the final state has one less electron, leading to the charging of the system. 
In the periodic boundary condition, a charged system cannot be treated in general because of the Coulombic divergence. 
The neutralization of the final state may occur in a metal, and theoretically such a neutralization can be 
justified as shown in Eq.~(4). However, it is unlikely that such a charge compensation takes place in an insulator
during the escape time of photoelectron ($\sim 10^{-16}$ sec.) \cite{Cavalieri2007}. 
To overcome the difficulty, here we propose a general method of treating the charged core hole state 
in the periodic boundary condition based on an exact Coulomb cutoff method \cite{Jarvis1997}. 
It is considered that the created core hole is isolated in the sample, resulting in violation of 
the periodicity of the system, while the charge density may recover from the perturbed one with distance from the core hole. 
The isolation of the core hole can be treated by dividing the charge density
$\rho_{\rm f}({\bf r})$ for the final state into a periodic part $\rho_{\rm i}({\bf r})$ and a non-periodic part 
$\Delta \rho({\bf r})(\equiv \rho_{\rm f}({\bf r})-\rho_{\rm i}({\bf r}))$ of which integration over the unit cell
is exactly a minus one, where $\rho_{\rm i}({\bf r})$ is the charge density for the initial state without the core hole. 
Then, as shown in Fig.~2(a) the Hartree potential $V_{\rm H}({\bf r})$ in the final state is given by 
\begin{eqnarray}
   V_{\rm H}({\bf r}) 
   = 
  V_{\rm H}^{({\rm P})}({\bf r})+
  V_{\rm H}^{({\rm NP})}({\bf r}),
\end{eqnarray}
where $V_{\rm H}^{({\rm P})}({\bf r})$ is the periodic Hartree potential calculated using the periodic part
$\rho_{\rm i}({\bf r})$ 
via a conventional method using a fast Fourier transform (FFT) for the Poisson equation.
On the other hand, the non-periodic Hartree potential $V_{\rm H}^{({\rm NP})}({\bf r})$ is calculated 
using $\Delta \rho({\bf r})$ and an exact Coulomb cutoff method by 
\begin{eqnarray}
  V_{\rm H}^{({\rm NP})}({\bf r}) = \sum_{\bf G} \tilde{\Delta \rho}({\bf G})\tilde{v}({\bf G}){\rm e}^{i{\bf G}\cdot{\bf r}},
\end{eqnarray}
where $\tilde{\Delta \rho}({\bf G})$ is the discrete Fourier transform of $\Delta \rho({\bf r})$, and $\tilde{v}({\bf G})$ is 
given by $\frac{4\pi}{G^2}(1-\cos(GR_{\rm c}))$, which is the Fourier transform of a cutoff Coulomb potential 
with the cutoff radius of $R_{\rm c}$ \cite{Jarvis1997}.
If $\Delta \rho({\bf r})$ is localized within a sphere of a radius $R$ as shown in Fig.~2(b), 
the extent of the Coulomb interaction is $2R$ at most in the sphere, which leads to $R_{\rm c}=2R$. 
In addition, a condition $2R<R'$ or equally $4R<L$ should be satisfied to avoid the spurious interaction 
between the core holes. In practice, we set $R_{\rm c}=\frac{1}{2}L$, and investigate the convergence of 
the binding energy as a function of $L$. With the treatment the core hole is electrostatically 
isolated from the other periodic images of the core hole even under the periodic boundary condition.

\begin{figure}[t]
    \centering
    \includegraphics[width=8.5cm]{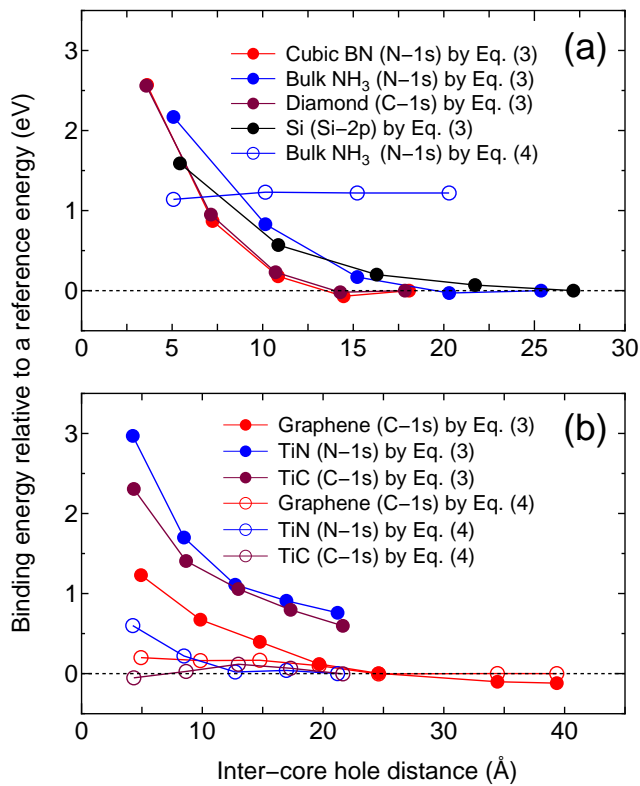}
    \caption{ 
     Calculated binding energies, relative to a reference energy, of (a) gapped systems 
     and (b) a semimetal (graphene) and metals as a function of inter-core hole distance.  
     The reference binding energies in (a) and (b) were calculated by Eqs.~(3) and (4), respectively, 
     for the largest unit cell for each system.
          }
\end{figure}

We implemented the method explained above in a DFT software package OpenMX \cite{OpenMX}, which is based on norm-conserving 
relativistic pseudopotentials \cite{Theurich2001,MBK1993} and pseudo-atomic basis functions \cite{Ozaki2003}. 
The pseudopotentials were generated including the $1s$-state for a carbon, nitrogen, and oxygen atom, and 
up to the $2p$-states for a silicon atom, respectively. 
As basis functions variationally optimized double valence plus single polarization orbitals (DVSP) 
and triple valence plus double polarization orbitals (TVDP) were used for bulks and gaseous molecules, 
respectively, after careful benchmark calculations for the convergence. A generalized gradient 
approximation (GGA) \cite{Perdew1996} to the exchange-correlation functional and an electronic 
temperature of 300~K were used. 
All the molecular and crystal structures we used in the study were taken from experimental ones.
Figures 3(a) and (b) show the relative binding energy of a core level in gapped systems and metals
including a semimetal (graphene), respectively, as a function of inter-core hole distance.
For the gapped systems the convergent results are obtained at the inter-core hole distance of 
$\sim$ 15, 20, and 27~\AA~for cubic boron nitride (diamond), bulk NH$_3$, and silicon, respectively. 
   \begin{table}[t]
     \caption{
        Calculated binding energy of a core level in bulks.
       }
   \vspace{1mm}
   \begin{tabular}{lccc}
   \hline\hline
      Material    & State  & Calc. (eV) & Expt. (eV)\\
     \hline
      {\it Gapped system}  & & &\\ 
      c-BN        & N-$1s$   &  398.87    & 398.1$^{*}$\\
      bulk NH$_3$ & N-$1s$   &  398.92    & 399.0$^{+}$\\
      Diamond     & C-$1s$   &  286.50    & 285.6$^{\dag}$\\
      Si          & Si-$2p_{1/2}$ & 100.13 & 99.8$^{*}$\\
      Si          & Si-$2p_{3/2}$ &  99.40 & 99.2$^{*}$\\
    \hline
      {\it Semimetal or Metal} & & &\\ 
      Graphene    & C-$1s$   &  284.23    & 284.4$^{\dag}$\\
      TiN         & N-$1s$   &  396.43    & 397.1$^{\S}$\\
      TiC         & C-$1s$   &  281.43    & 281.5$^{*}$\\
   \hline
   \end{tabular}
   \\
   $^*$ Ref.~\cite{Perkin1979}, $^+$ Ref.~\cite{Larkins1979}, $^\dag$ Ref.~\cite{Yan2004} (graphite), $^\S$ Ref.~\cite{Jaeger2013}
  \end{table}
This implies that the difference charge $\Delta \rho({\bf r})$ induced by the creation of the core hole
is localized within a sphere with a radius of $R=L/4$, e.g., $\sim$ 7~\AA~for silicon.
In fact, the localization of $\Delta \rho({\bf r})$ in silicon can be confirmed 
by the distribution in real space and the radial distribution of a spherically averaged $\Delta \rho$
as shown in Figs.~4(a) and (b). The deficiency of electron around 0.3~\AA~corresponding to 
the core hole in the $2p$-states is compensated by the increase of electron density around 1~\AA,
which is the screening on the same silicon atom for the core hole. 
As a result of the short range screening, the non-periodic Hartree potential $V_{\rm H}^{({\rm NP})}({\bf r})$ 
deviates largely from $-1/r$ as shown in Fig.~4(c). 
In Fig.~3(a) it is also shown that the binding energy of the bulk NH$_3$ calculated 
with Eq.~(4) converges at 1.2 eV above, implying that Eq.~(4) cannot be applied to 
the gapped system.
On the other hand, for the metallic cases it is found that Eq.~(4) provides 
a much faster convergence than Eq.~(3), and both Eqs.~(3) and (4) seem to give 
a practically equivalent binding energy, while the results calculated with Eq.~(3) for 
TiN and TiC do not reach to the sufficient convergence due to computational limitation \cite{gra_comp}. 
We further verified the equivalence between Eqs.~(3) and (4) for metals by calculating 
the binding energy of the $1s$-state of a carbon atom in a model metallic system of an infinite carbon 
chain with the nearest neighbor distance of 1.5~\AA, which is computationally accessible, 
and found that the difference between the two values is 0.07 eV at the inter-core hole distance 
of 225~\AA~(not shown in Fig.~3(b)). 
\begin{figure}[tb]
 \begin{minipage}{17.5cm}
    \centering
    \includegraphics[width=16cm]{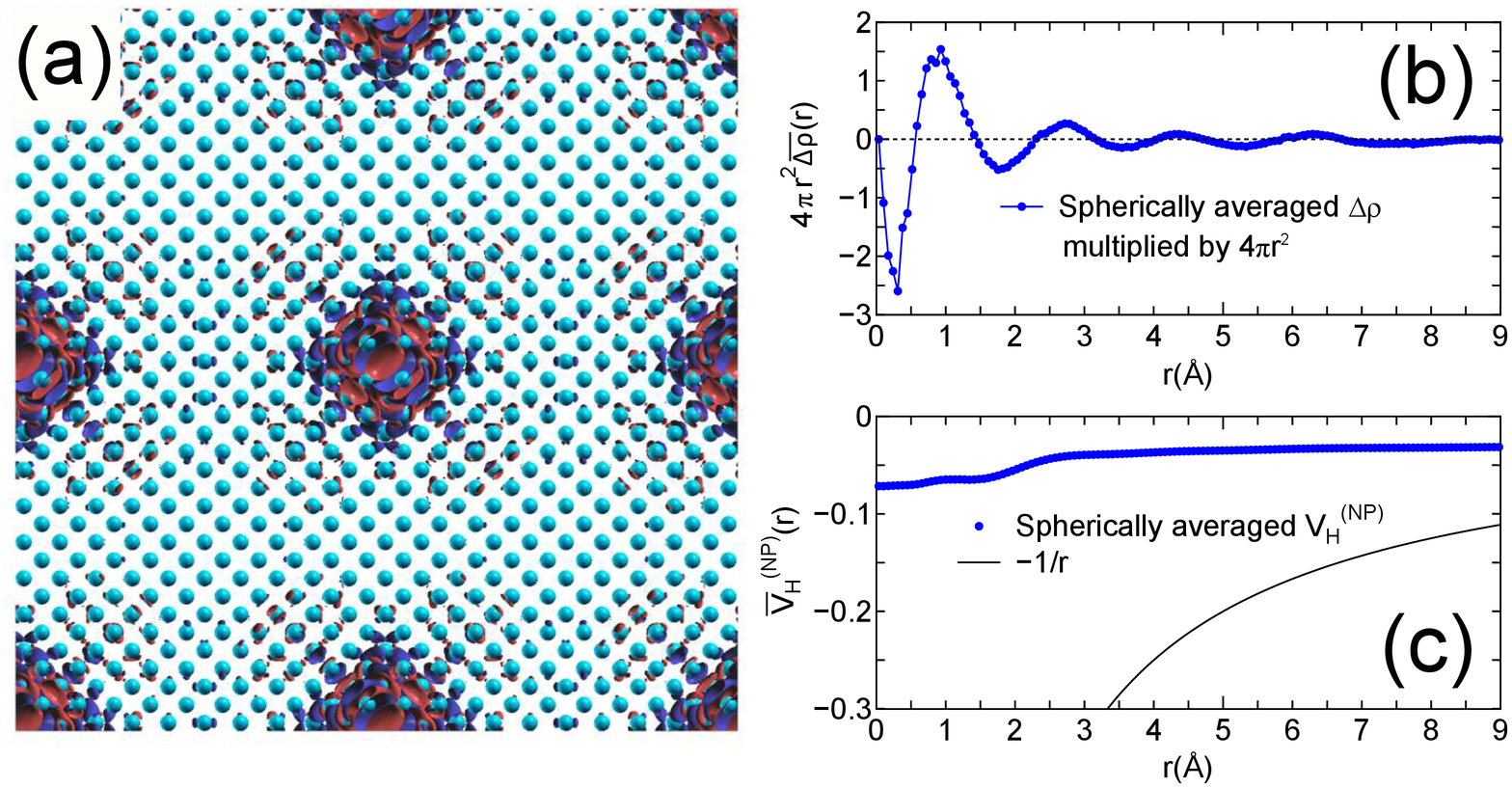}
    \caption{
             (a) Difference charge density $\Delta \rho$
             in silicon, induced by the creation of a core hole in the $2p$-states,
             where the unit cell contains 1000 atoms, and the inter-core hole distance is 27.15~\AA. 
             (b) Radial distribution of $4\pi r^2 \overline{\Delta \rho}$, where
             $\overline{\Delta \rho}$ is a spherically averaged $\Delta \rho$.
             (c) Radial distribution of $\overline{V}_{\rm H}^{({\rm NP})}$ being 
             a spherically averaged $V_{\rm H}^{({\rm NP})}$. 
             }
 \end{minipage}
\end{figure}
However, the convergence with Eq.~(3) is again found to be very slow.
Therefore, Eq.~(4) is considered to be the choice for 
the practical calculation of a metallic system because of the faster convergence.
By compiling the size of the unit cell achieving the convergence into the number of atoms 
in the unit cell, the use of a supercell including $\sim$ 500 and 64 atoms 
for gapped and metallic systems in three-dimensions might be a practical guideline for achieving 
a sufficient convergence by using Eqs.~(3) and (4), respectively.
The calculated values of binding energies are well compared with the experimental 
absolute values as shown in Table I for both the gapped and metallic systems, 
and the mean absolute (percentage) error is found to be 0.4 eV (0.16~\%) for the eight cases.
We see that the splitting due to spin-orbit coupling in the silicon $2p$-states is 
well reproduced. 
In addition, binding energies of a core level for gaseous molecules calculated by Eq.~(5)
are shown in the supplemental material, where the mean absolute (percentage) error is found to be 
0.5 eV (0.22~\%) for the 23 cases \cite{ex_sp}.

\begin{figure}[th]
  \centering
  \includegraphics[width=8.5cm]{blank.eps}
\end{figure}

In summary, we proposed a general method to calculate absolute binding energies of core levels
in metals and insulators in a framework of DFT. The method is based on a penalty functional 
and an exact Coulomb cutoff method. The former allows us to calculate multiplet splittings 
due to the chemical shift, spin-orbit coupling, and exchange interaction, and to perform geometry optimization 
of a system with a core hole if necessary, while the latter enables us to treat a charged 
system with a core hole under the periodic boundary condition. 
It was also shown that especially for metals Eq.~(4) involving the neutralized 
final state is equivalent to Eq.~(3) involving the ionized final state, and 
that Eq.~(4) is computationally more efficient than Eq.~(3). 
The series of calculations and the good agreement with the binding energies
measured in the XPS experiments clearly demonstrate that the proposed method is 
general and accurate for a variety of materials. 
Considering the importance of the informative XPS measurement in materials researches, 
it is anticipated that the proposed method will be an indispensable theoretical tool 
to quantitatively analyze the absolute binding energy of core levels 
in both metals and insulators.

We would like to thank Y. Yamada-Takamura and J. Yoshinobu for helpful discussions
on the XPS measurements.
This work was supported by Priority Issue (Creation of new functional 
devices and high-performance materials to support next-generation 
industries) to be tackled by using Post 'K' Computer, MEXT, Japan.

\newpage

\noindent
{\bf\Large Supplemental material}\\[3mm]

   \begin{table}[h]
     \begin{flushleft}
    TABLE S-I.
      Calculated binding energies of core levels in gaseous systems. 
      The Si-$2p$ states were calculate by a scalar relativistic treatment \cite{Koelling1977}.
      The non-equivalent nitrogen atoms in a N$_2$O molecule are specified by the underline. 
      \\[2mm]
     \end{flushleft}

   \vspace{1mm}
   \begin{tabular}{lccc}
   \hline\hline
      Molecule    & Calc. (eV) &  & Expt.$^{*}$ (eV)\\
     \hline
      {\it C-1s state} & & &\\ 
      CO          &  295.87  & \hspace{3mm} & 296.19\\
      C$_2$H$_2$  &  291.24  & \hspace{3mm} & 291.17\\
      CO$_2$      &  296.89  & \hspace{3mm} & 297.66\\
      HCN         &  293.35  & \hspace{3mm} & 293.50\\
      C$_2$H$_4$  &  290.50  & \hspace{3mm} & 290.79\\
      H$_2$CO     &  294.00  & \hspace{3mm} & 294.47\\
    \hline
      {\it N-1s state} & & &\\ 
      N$_2$       &  409.89  & \hspace{3mm} & 409.83\\
      NH$_3$      &  404.70  & \hspace{3mm} & 405.60\\
      N$_2$H$_4$  &  404.82  & \hspace{3mm} & 406.1\\
      HCN         &  406.16  & \hspace{3mm} & 406.36\\
      \underline{N}NO
                  &  408.24  & \hspace{3mm} & 408.66\\
      N\underline{N}O
                  &  411.98  & \hspace{3mm} & 412.57\\
      NO(S=0)     &  410.62  & \hspace{3mm} & 411.6 \\
      NO(S=1)     &  410.10  & \hspace{3mm} & 410.2 \\
    \hline
      {\it O-1s state} & & &\\ 
      CO          &  542.50  & \hspace{3mm} & 542.4\\
      CO$_2$      &  541.08  & \hspace{3mm} & 541.2\\
      O$_2$(S=$\frac{1}{2}$) &  543.15  & \hspace{3mm} & 544.2\\
      O$_2$(S=$\frac{3}{2}$) &  542.64  & \hspace{3mm} & 543.1\\
      H$_2$O      &  539.18  & \hspace{3mm} & 539.9\\
    \hline
      {\it Si-2p state} & &\\ 
      SiH$_4$     &  106.56  & \hspace{3mm} & 107.3\\
      Si$_2$H$_6$ &  106.21  & \hspace{3mm} & 106.86\\
      SiF$_4$     &  111.02  & \hspace{3mm} & 111.7\\
      SiCl$_4$    &  109.32  & \hspace{3mm} & 110.2\\
   \hline
   $^*$ \cite{Bakke1980,Jolly1984,Morioka1974,Wight1974}
   \end{tabular}
  \end{table}


\begin{thebibliography}{99}

% Siegbahn's book
\bibitem{Siegbahn1967}
K. Siegbahn, C. Nordling, A. Fahlman, K. Hamrin, J. Hedman, R. Nordberg,
C. Johansson, T. Bergmark, S.-E. Karlsson, I. Lindgren, and B. Lindberg, 
{\it Atomic, molecular and solid-state structure studied by means of electron spectroscopy},
Nova Acta Regiae Soc. Sci. Ups. 20.1-282, Almqvist and Wiksells (1967).

% XPS measurement for gaseous systems
\bibitem{Siegbahn1969}
K. Siegbahn, C. Nordling, G. Johansson, J. Hedman, P.-F. Heden, K. Hamrin,
U. Gelius, T. Bergmark, L.O. Werme, R. Manne, and Y. Baer, 
{\it ESCA Applied to Free Molecules}, North-Holland, Amsterdam, The Netherlands (1969).

% Review of XPS
\bibitem{Fadley2010}
C.S. Fadley, J. Electron Spectrosc. Relat. Phenom. {\bf 178-179}, 2 (2010). 

% core level satellites
\bibitem{Bagus2010}
P.S. Bagus, C.J. Nelin, E.S. Ilton, M. Baron, H. Abbott, E. Primorac, H. Kuhlenbeck, S. Shaikhutdinov, and H.-J. Freund,
Chem. Phys. Lett. {\bf 487}, 237 (2010). 

% core level vibrational fine structure
\bibitem{Hergenhahn2004}
U. Hergenhahn, J. Phys. B: Atom. Mol. Opt. Phys. {\bf 37}, R89 (2004).

% magnetic circular dichroism (MCD)
\bibitem{Yang2002}
S.-H. Yang, B.S. Mun, N. Mannella, S.-K. Kim, J.B. Kortright, J. Underwood, F. Salmassi, 
E. Arenholz, A. Young, Z. Hussain, M.A. Van Hove, and C.S. Fadley, 
J. Phys. Condens. Matter {\bf 14}, L406 (2002).

% spin-resolved XPS
\bibitem{Kim2001}
H.-J. Kim, E. Vescovo, S. Heinze, and S. Bl\"{u}gel, Surf. Sci. {\bf 478}, 193 (2001).

% photoelectron holography
\bibitem{Omori2002}
S. Omori, Y. Nihei, E. Rotenberg, J.D. Denlinger, S. Marchesini, S.D. Kevan, 
B.P. Tonner, M.A. Van Hove, and C.S. Fadley, Phys. Rev. Lett. {\bf 88}, 055504 (2002).

% the first interpretation of the photoemission 
\bibitem{Einstein1905}
A. Einstein, Ann. Phys. 17, {\bf 132} (1905).

% screened core hole pseudopotential
\bibitem{Pehlke1993}
E. Pehlke and M. Scheffler, Phys. Rev. Lett. {\bf 71}, 2338 (1993). 

% adding an excess charge into the conduction bands
\bibitem{Susi2015}
T. Susi, D.J. Mowbray, M.P. Ljungberg, and P. Ayala, 
Phys. Rev. B {\bf 91} 081401(R) (2015).

\bibitem{Olovsson2010}
W. Olovsson, T. Marten, E. Holmstr\"{o}m, B. Johanssone, and I.A. Abrikosov,
J. Electron Spectrosc. Relat. Phenom. {\bf 178-179}, 88 (2010). 

% cluster models
\bibitem{Bagus2013}
P.S. Bagus, E.S. Ilton, and C.J. Nelin,  
Surf. Sci. Reports {\bf 68}, 273 (2013).

% the escape time 
\bibitem{Cavalieri2007}
A.L. Cavalieri, N.M\"{u}lle, Th. Uphue, V S. Yakovle, A. Balt\v{u}ˇk, B. Horvat, B. Schmid, L. B\"{u}mel,
R. Holzwarth, S. Hendel, M. Drescher, U. Kleineberg, P.M. Echenique, R. Kienberger, F. Krausz, and U. Heinzmann,
Nature {\bf 449}, 1029 (2007).

% Slater transition state theory
\bibitem{Slater1974}
J.C. Slater, {\it The Self-Consistent Field for Molecules and Solids, Quantum Theory of Molecules and Solids} 
(McGraw-Hill, New York, 1974), Vol. 4.

% DFT
\bibitem{Hohenberg1964}
P. Hohenberg and W. Kohn, Phys. Rev. {\bf 136}, B864 (1964).

\bibitem{Kohn1965}
W. Kohn and L. J. Sham, Phys. Rev. {\bf 140}, A1133 (1965). 

% recoil energy
\bibitem{Watts1994}
J.F. Watts, Vacuum {\bf 45}, 653 (1994). 

% calibation by the Fermi edge of silver
\bibitem{King2007}
P.D.C. King, T.D. Veal, P.H. Jefferson, C.F. McConville, H. Lu, and W.J. Schaff,
Phys. Rev. B {\bf 75}, 115312 (2007).

% Janak's theorem
\bibitem{Janak1978}
J.F. Janak,
Phys. Rev. B {\bf 18}, 7165 (1978). 

% exact Coulomb cutoff method
\bibitem{Jarvis1997}
M.R. Jarvis, I.D. White, R.W. Godby, and M.C. Payne, Phys. Rev. B {\bf 56}, 14972 (1997).

% OpenMX
\bibitem{OpenMX}
The code, OpenMX, pseudo-atomic basis functions, and pseudopotentials are
available on a web site (http://www.openmx-square.org/).

% MBK pseudopotentials
\bibitem{Theurich2001}
G. Theurich and N.A. Hill, Phys. Rev. B {\bf 64}, 073106 (2001).

\bibitem{MBK1993}
I. Morrison, D.M. Bylander, L. Kleinman, Phys. Rev. B {\bf 47}, 6728 (1993).

% OpenMX basis functions
\bibitem{Ozaki2003}
T. Ozaki, Phys. Rev. B. {\bf 67}, 155108, (2003). 

% GGA-PBE
\bibitem{Perdew1996}
J.P. Perdew, K. Burke, and M. Ernzerhof, Phys. Rev. Lett. {\bf 77}, 3865 (1996).

% Handbook of XPS, c-BN, 
\bibitem{Perkin1979}
C.D. Wagner, W.M. Riggs, L.E. Davis, J.F. Monider, G.E. Mullenberg, 
"Handbook of X-ray photoelectron spectroscopy", Perkin-Elmer (1979).

% bulk NH3
\bibitem{Larkins1979}
F.P. Larkins and A. Lubenfeld, 
J. Electron Spectrosc. Relat. Phenom. {\bf 15}, 137 (1979).

% diamond
\bibitem{Yan2004}
X.B. Yan, T. Xu, S.R. Yang, H.W. Liu, and Q.J. Xue,
J. Phys. D: Appl. Phys. {\bf 37}, 2416 (2004).

% TiN
\bibitem{Jaeger2013}
D. Jaeger and J. Patscheider,
Surf. Sci. Spectra {\bf 20}, 1 (2013).

% Eqs. (3) and (4) for graphene
\bibitem{gra_comp}
It is noted that the convergent values calculated with Eqs.~(3) and (4) deviate slightly 
from each other for graphene, suggesting that Eq.~(4) is not equivalent to Eq.~(3) 
for a semi-metal.  

% on exchange-splitting 
\bibitem{ex_sp}
The underestimation of splittings in a O$_2$ and NO molecule should be attributed to 
an insufficient treatment of the exchange interaction in the semi-local functional 
rather than the proposed method, since the Hartree-Fock method accurately reproduces 
the splitting \cite{Bagus1971}, suggesting the necessity of non-local functionals 
for an accurate description of the intra-atomic exchange interaction.

% exchange-splitting of NO
\bibitem{Bagus1971}
P.S. Bagus and H.F. Schaefer III,
J. Chem. Phys. {\bf 55}, 1474 (1971).


\end{thebibliography}

\begin{thebibliography}{99}

% gaseous systems
\bibitem{Bakke1980}
A.A. Bakke, A.W. Chen, and W.L. Jolly, 
J. Electron Spectrosc. Relat. Phenom. {\bf 20}, 333 (1980).

% gaseous systems
\bibitem{Jolly1984}
W.L. Jolly, K.D. Bomben, C.J. Eyermann, 
At. Data Nul. Data Tables {\bf 31}, 433 (1984).

% NO
\bibitem{Morioka1974}
Y. Morioka, M. Nakamura, E. Ishiguro, and M. Sasanuma, J. Chem. Phys. {\bf 61}, 1426 (1974).

% O2
\bibitem{Wight1974}
G.R. Wight and C.E. Brion, J. Electron Spectrosc. Relat. Phenom. {\bf 4}, 313 (1974).

% scalar relativistic 
\bibitem{Koelling1977}
D. D. Koelling and B. N. Harmon, J. Phys. C: Solid State Phys. {\bf 10}, 3107 (1977)

\end{thebibliography}
\end{document}